\documentclass[a4paper, 10 pt, conference]{ieeeconf}  

\IEEEoverridecommandlockouts                              

\usepackage{graphics} 
\usepackage{epsfig} 
\usepackage{mathptmx} 
\usepackage{times} 
\usepackage{amsmath} 
\usepackage{amssymb}  
\usepackage[T1]{fontenc}
\usepackage{amsfonts}
\usepackage{booktabs}
\usepackage{siunitx}
\usepackage{caption} 
\usepackage{algorithm}
\usepackage{algorithmic}
\usepackage[flushleft]{threeparttable}
\title{\LARGE \bf
Integrative Object and Pose to Task Detection for an Augmented-Reality-based Human Assistance System using Neural Networks
}

\author{Linh K\"astner$^{1}$, Leon Eversberg$^{1}$, Marina Mursa$^{1}$ and Jens Lambrecht$^{1}$
\thanks{$^{1}$Linh K\"astner, Leon Eversberg, Marina Mursa and Jens Lambrecht are with the Chair Industry Grade Networks and Clouds, Faculty of Electrical Engineering, and Computer Science,
       Berlin Institute of Technology, Berlin, Germany
        {\tt\small linhdoan@tu-berlin.de}}%
}

\begin{document}

\maketitle
\thispagestyle{empty}
\pagestyle{empty}

\begin{abstract}
As a result of an increasingly automatized and digitized industry, processes are becoming more complex. Augmented Reality has shown considerable potential in assisting workers with complex tasks by enhancing user understanding and experience with spatial information. However, the acceptance and integration of AR into industrial processes is still limited due to the lack of established methods and tedious integration efforts. Meanwhile, deep neural networks have achieved remarkable results in computer vision tasks and bear great prospects to enrich Augmented Reality applications .
In this paper, we propose an Augmented-Reality-based human assistance system to assist workers in complex manual tasks where we incorporate deep neural networks for computer vision tasks. More specifically, we combine Augmented Reality with object and action detectors to make workflows more intuitive and flexible. To evaluate our system in terms of user acceptance and efficiency, we conducted several user studies. We found a significant reduction in time to task completion in untrained workers and a decrease in error rate. Furthermore, we investigated the users learning curve with our assistance system.
\end{abstract}

\section{Introduction}
\noindent Due to the raise of automation and digitization in all areas of industries, the complexity of manufacturing processes increased in recent years and tools to handle the rapidly growing amounts of information are needed \cite{Preuveneers2017intelligent}.
Despite the upcoming changes in the industrial workforce, it is unlikely that human labor will become obsolete due to complete automation. Manufacturers look for ways to use technological advancement to assist their employees cognitively and physically rather than replacing them entirely \cite{manandmachine}.
These assistance systems can help with the global trend of an increasing demand in high-qualification- and decreasing demand in low-qualification-jobs \cite{cedefop-forecast}. In this context, Augmented-Reality (AR) -based assistance systems have shown promising results such as reducing task execution time, improving product quality or increasing workers ability to learn new tasks\,\cite{SouzaCardoso2020}. While AR assistance systems have great potential of improving the collaboration between humans and machines, the implementation of these systems face technological, environmental and organisational challenges\,\cite{Masood2019} and are still an ongoing field of research. Furthermore, state-of-the-art AR assistance systems often require an additional comissioning step, are restricted to specific areas and do not take advantage of the recent success of computer vision approaches like neural networks \cite{bosch-hololens, uva, Chu2020Comparing}. The majority of AR systems still rely on fiducial markers \cite{SouzaCardoso2020, Bottani2019Augmented} or manual calibration  \cite{bosch-hololens, uva, Chu2020Comparing}. Additionally, while most AR systems have integrated some form of object detection, the addition of an action detection is not yet established. Moreover, only little attention has been put on static screens as AR visualization hardware\,\cite{Egger2020Augmented} despite their simplicity and potential to employ more processing power compared to handheld devices or head-mounted devices (HMDs).

\begin{figure}[]
	\centering
	\includegraphics[width=3.3in, height=2.4in]{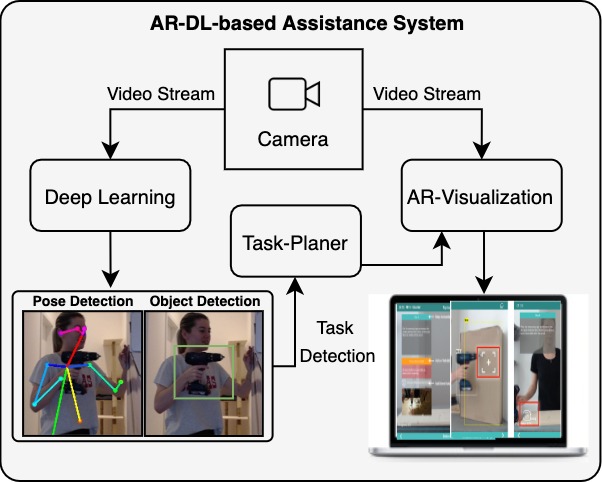}
	\caption{Proposed pipeline of industrial AR assistance system}
	\label{intro}
\end{figure}

On this account, we present an AR-based industrial assistance system, which incorporates deep learning (DL) architectures to ensure a more flexible and intuitive user experience. We combine an object detection model with an action detection model to guide untrained users in manual tasks with the objective of promoting knowledge-transfer (s. Fig. \ref{intro}). We tested our assistance system with 30 participants in an exemplary assembly scenario consisting of 9 steps and evaluated the results in terms of task completion time, error rate, user acceptance and learning promotion.
The main contributions of this paper are the following: 
\begin{itemize}
		\item A proposal of an industrial AR assistance system for complex manual tasks using deep neural networks relying on RGB camera input only
		\item The combination of object detection and action detection enabling the system to perform interaction analysis
		\item A quantitative and qualitative evaluation of our system such as time to task completion, error rate and a user acceptance questionnaire
		\item A novel study on the effect of task repetition on the impact of the AR assistance system
	\end{itemize}

\noindent The paper is structured as follows: Sec. II gives an overview of related work. Sec. III presents the methodology. Sec. IV presents the results and evaluation. Finally, Sec. V gives a conclusion of our work.

\section{RELATED WORK}
\noindent Over the last decade there has been a growing interest in industrial AR. The term Augmented Reality was first introduced by Caudell and Mizell \cite{Caudell1992Augmented} in 1992 where they used HMDs in aircraft manufacturing to reduce manufacturing cost and improve workers efficiency. The largest focus of research has been in the application of manual assembly tasks\,\cite{SouzaCardoso2020}. 
\begin{figure*}[!h]
	\centering
	\includegraphics[width=6.7in]{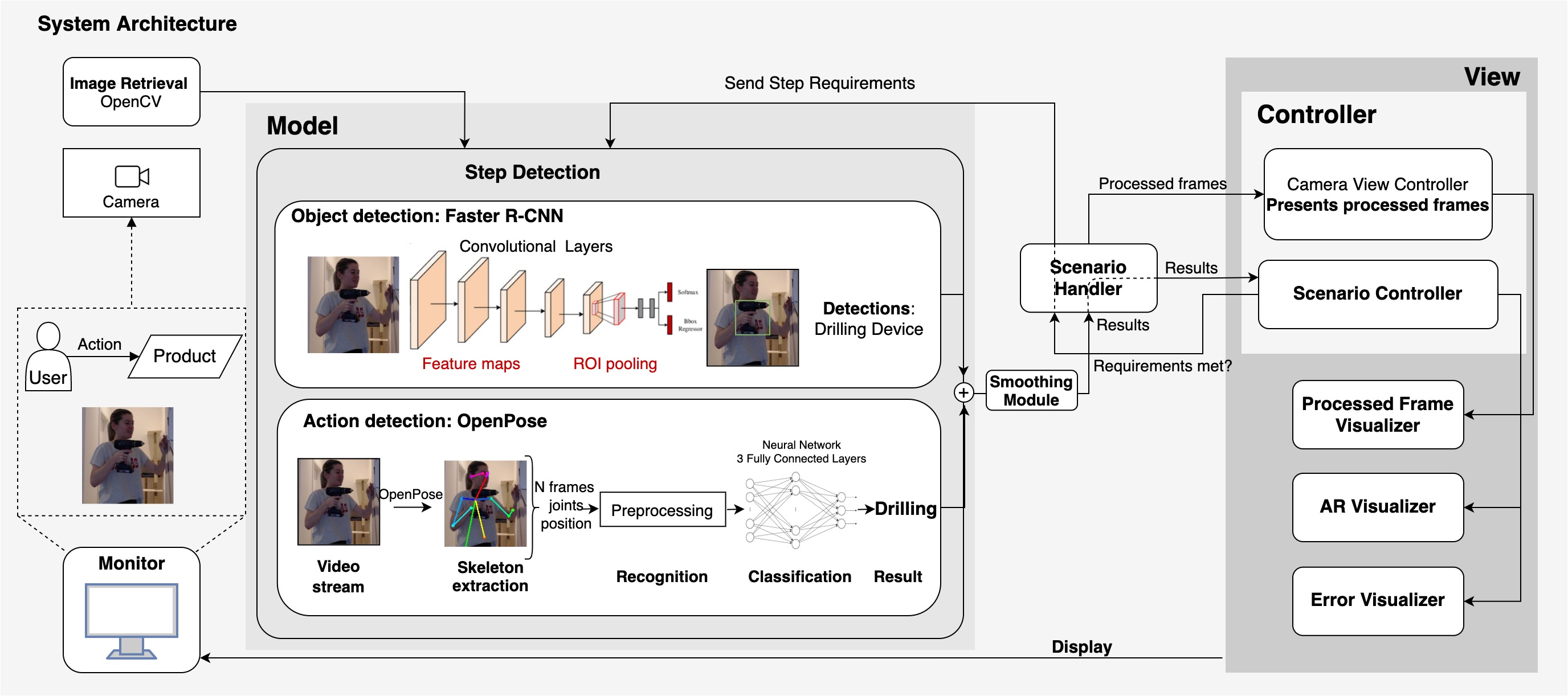}
	\caption{System design of our proposed assistance system}
	\label{arch2}
\end{figure*}
\noindent 
Loch  et  al. \cite{Loch2016Comparing} created an AR assistance system for manual assembly tasks of Lego bricks. They provide assistance on a screen by overlaying animations over the video feed of a camera. They compared their AR system to video instructions and found an improvement in time to completion and number of errors in participants when using the AR system for the first time. Uva  et  al. \cite{uva} designed a Spatial Augmented Reality (SAR) prototype intended to be used for working stations in the context of smart factories. They used a workbench consisting of a controller, a projector and camera mounted on a frame and a rotating table with fiducial markers. The instructions were projected directly on the objects to be maintained. Their research shows that SAR instructions have significant benefits compared to paper-based instructions in tasks with high difficulty. The authors note that they did not consider the participants experience or the learning effect after multiple task repetitions. In contrast to  \cite{uva}, Funk et al. \cite{Funk2017Working} did a long-term analysis of their SAR assistance system. The researchers equipped an assembly line of a car manufacturing company with projectors and depth cameras and tested their prototype over 11 work days with expert and untrained workers. The depth images were used to detect the picking of boxes in a specific location and direct feedback was given by a top-mounted projector. Subsequently, the task completion time, error rate and NASA-TLX scores were compared and found no improvements when comparing in-situ instructions with no instructions. Instead, it was observed that the assistance system was useful for untrained workers during the learning phase and the researchers suggest that there is a tipping point after which the assistance system loses its benefits, thus suggesting that AR assistance systems are best for knowledge transfer and helping untrained workers to learn new workflows. Similar to our work, Chu et al. \cite{Chu2020Comparing} recently published the full framework of an AR assistance system to assemble Dougong structures. Their AR smartphone app uses object recognition to identify relevant parts of the current assembly step and then helps with the assembly process by displaying AR animations. Object recognition and tracking is handled by the AR library \textit{Vuforia} and the display of assembly instructions are implemented with \textit{Unity}. 

\noindent While most studies of AR assistance systems are done in a laboratory environment \cite{Egger2020Augmented}, Lorenz et al. \cite{Lorenz2018Industrial} analyzed requirements of three different production environments in the real world and conclude that rugged tablet computers are the best hardware choice because of harsh production environments and the worn safety gear of the workers. The researchers point out that HMDs, which are the most researched AR hardware \cite{Egger2020Augmented}, are neither robust nor can they be worn with protective gear. Additionally, Wolfartsberger et al. \cite{Wolfartsberger2019Perspectives} also noted that the current generation of HMDs still has many disadvantages and recommend using a fixed tablet screen or in-situ projection.

\noindent In comparison to the state of the art on AR assistance systems, our work is based on deep neural networks for object detection and human pose detection. The addition of our action detection system enables the AR system to analyze the workers actions and incorporates the trending field of human-object interaction into assistance systems. In contrast to other works \cite{Loch2016Comparing, uva, Funk2017Working}, we describe the full framework of our system architecture. Because our machine-learning-based system only relies on RGB video input and a monitor for display, our assistance system is very flexible and complies with the compiled requirements of \cite{Lorenz2018Industrial} for harsh industrial environments.

\section{Methodology}
\noindent Our objective is to provide an intuitive system that aims to increase the efficiency of the user by using the following integrated features: detailed process step descriptions, visual hints using AR visualizations and video hints by using video tutorials.
\noindent All these features make sure to decrease the time spent of looking for additional sources of information, which potentially interrupts the workers focus, thus decreasing his efficiency. 
Another objective is the prevention of errors or the timely detection of errors such that the user can instantly react.
The assistance system aims to reduce the error rate by the use of the following implemented features:
action error detection, object identification, step validation and error prevention - through the use of detailed instructions, hints and guiding AR elements.

\subsection{System Architecture}
\noindent The overall architecture of our proposed assistance system can be observed in Fig. \ref{arch2}. It offers a high-level view of the system and highlights all entities.
\noindent As illustrated in the system design, the main business logic is divided between a \textit{Step Detection} and a \textit{Scenario Handler}. The \textit{Step Detection} consumes the camera input in form of frames. The \textit{Scenario Handler} provides the step requirements such as necessary objects and actions the worker has to do at the respective steps. These information are sent to the object and action detection models, which are running in parallel in order to increase the speed of image processing and therefore assure faster responsiveness. We used a \textit{Faster R-CNN} architecture \cite{FasterRCNN} for object detection with an \textit{Inception v2} model \cite{InceptionV2}, while \textit{OpenPose} \cite{Cao2019OpenPose} together with a neural network based frame classifier are utilized for the action detection. After the models process the frame, the results are sent as a tuple to a queue which the \textit{Camera View Controller} pulls and displays to the user. The \textit{Scenario Controller} is responsible for evaluating these results in order to display possible errors or wrong actions. Furthermore, it is also responsible for displaying AR components to assist the user in fixing these errors and giving supplementary hints. Additionally, in order to improve the decision making accuracy and avoid flickering of data, a smoothing module is employed which runs over the last 10 frames and takes the result with the highest occurrence to be considered truth. 
\noindent In case the step was successfully validated, the \textit{Scenario Controller} sends a validation event back to the \textit{Scenario Handler} which moves to the next step. Once a new step is initiated, the \textit{Scenario Controller} is notified by the event and displays it in the description of the new step. \noindent In case a step was not validated, an error event will be published which notifies the \textit{Scenario Handler} to display an error/warning message through the \textit{Error View}.

\subsection{Action Detection}

\noindent In order to recognize the workers actions and thereby detect the correctness of the users action, we use the skeleton-tracking-based pose detection of \textit{OpenPose} over a series of frames together with a neural network classifier in our assistance system. The implementation is based on the open source project \textit{Realtime-Action-Recognition} by Chen\,et\,al\,\cite{githubchen}. First, we calculate the joint velocities from a series of N=5 frames using the formulas described in equation (1) to (4). Afterwards, the actions are derived by a neural network which outputs the classified action. 
To retrieve the joints of each frame, the \textit{OpenPose} algorithm is trained to detect the human skeleton from an image. The input to \textit{OpenPose} is an RGB image acquired from the camera and the output is the skeleton of the human. 

\noindent The raw skeleton data output of \textit{OpenPose} is pre-processed with the following three steps that are being proposed by\,\cite{githubchen} and applied to our use case:\\

\begin{enumerate}
	\item \textbf{Coordinate scaling} - As OpenPose has a different unit for \verb|x| coordinate and \verb|y| coordinate, the output joint's position must be scaled.
	
	    \item \textbf{Removal of unnecessary joints} - Mainly, the movements of arms and hands are relevant for our use case, thus the joints for the head are removed to reduce the vector size. 
	
	\item \textbf{Padding of missing joints}
	- To enhance robustness against inconsistent detections, we incorporate a padding with detections of the previous frames in case joints are missing.
	
\end{enumerate}


\noindent The joint positions are retrieved and joint positions from N=5 consecutive frames are considered to calculate the joint velocities and derive the user action. First, the height of the body $H$ is calculated to find the average coordinates \(x_1, y_1\) between the two hips that are represented by the coordinated pairs:  left hip: \((x_{11}, y_{11})\) and right hip: \((x_{12}, y_{12})\).

\begin{align}
      x_1, y_1 =  \frac{x_{11} + x_{12}}{2}, \frac{(y_{11} + y_{12})}{2} \\
        H = \sqrt {(x_0 - x_1)^2 + (y_0 - y_1)^2}
\end{align}
 Subsequently, the normalized joint positions \(X_n\) can be calculated using the initial joint coordinates \(X\). Finally, the joint velocity $V_i$ for all joints can be calculated where \(i\) represents the list of all joints and \([t_k]\) is the frame number from the series of the 5 frames.
\begin{align}
        X_{n} =  \frac{X}{H} \\
    	V_{i} = {X_n[t_k]-X_n[t_{k-1}]}
\end{align}
These features are concatenated and given as input into a 3 layer fully connected neural network to classify the users action.

\subsection{Graphical User Interface}
\begin{figure}[]
	\centering
	\includegraphics[width=3.3in, height=2.7in]{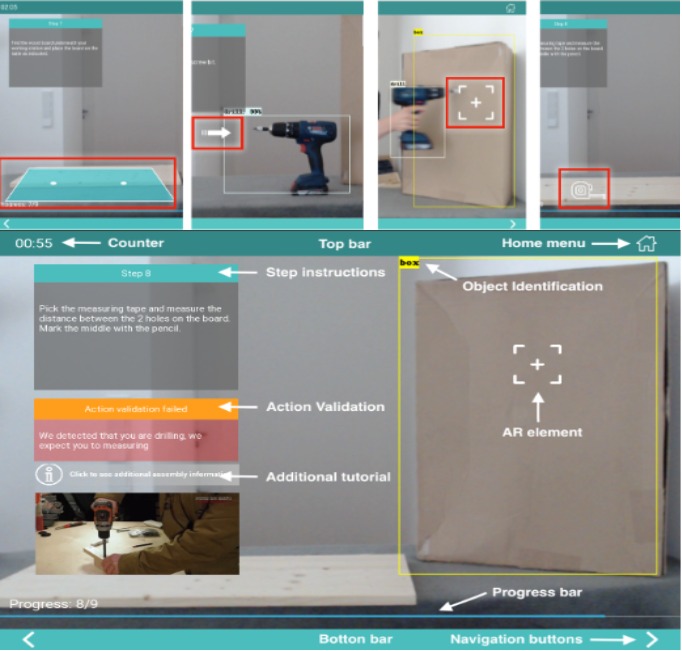}
	\caption{Graphical user interface with AR components (top) text and video for additional information as well as warnings or error messages (bottom)}
	\label{gui}
\end{figure}

\noindent Fig. \ref{gui} illustrates the most important features of our Graphical User Interface (GUI), which is implemented as a cross platform application using Kivy. It includes AR components such as virtual hints of relevant areas within the working place and animations that describe the task visually. Text fields are displayed to inform the user about tasks or errors. Additionally, videos can be displayed, which provide additional information. The progress bar at the bottom displays the step a worker is currently in and the watch counter at the top left shows the time passed in order to provide a reference to the user.
To display the AR components and animations, we used an anchor based calibration, which means that we utilized the position of a detected object as an anchor to displays the AR features with respect to that object. This way, we ensure the flexibility of our system by not relying on manually aligned positions. We employ a database where the specific display positions of AR components with respect to the objects are listed accordingly. 
\subsection{Experiments}
\noindent In order to evaluate the impact of our system on the worker, we conducted a series of user studies. For that reason, a special assembly scenario consisting of 9 steps was created which reflects all the features of the system such as: object detection, action validation, interface enhancement through AR elements and additional hints through video tutorials. Depending on the scope of the user study, certain features were activated or deactivated in order to study their impact. To create a comparable baseline, we provided a basic version of the system where we disabled any additional features and only kept descriptions of the 9 steps. Subsequently, we compared this baseline against our assistance system with all assistance features. Table \ref{tablehyper} shows the instructions of the test scenario. A total of 30 participants (13 female, 17 male) from various backgrounds were recruited. The participants were aged from 20 to 
45. 17 out of the 30 participants had a technical background. The rest had a business, medical, law or construction related background.

\begin{table}[!h]
	
	\renewcommand{\arraystretch}{1.3}
	\caption{Task instructions}
	\begin{tabular}{cp{6.5cm}}
		\hline
		Step number  & Explanation    \\ \hline
1 & Please make sure to have all required tools available on your work table. \\ 
\hline
2 & Pick up the drill and make a whole around 3 cm deep on the board at $x=10cm$, $y=15cm$, where the short side of the board is considered the x-axis and the longer side, the y-axis.  \\
\hline
3 & Find the screw bit inside the grey box and change the drill bit to a screw bit. \\
\hline
4 & Find the green box and pick a screw.\\
\hline
5 & Secure the screw with the drill into the previously made hole. \\
\hline
6 & Find the wooden board underneath your working station and place the board on the table. \\
\hline
7 & Pick the pencil and mark two spots with the pencil on the following coordinates on the board: 1) $x=5cm$, $y=20cm$ and 2) $x=15cm$, $y=20cm$. \\
\hline
8 &  Then, measure the distance between the two holes and mark the middle point. \\
\hline
9 & Pick up the hacksaw and saw the board in 2 based on the previously made middle mark. \\
\hline
	\hline
	\end{tabular}
	
	\label{tablehyper}
	
\end{table}
\section{Evaluation}
 \noindent To test whether the assumptions from the conceptual design were correct and if all the features led to an improved time of completion the task, we measured the task completion time as well as error rate in two modes as described above. In the first mode, denoted as mode 1, all features were removed except for \emph{the step instructional text} and the users were asked to complete the scenario. In the second mode, denoted as mode 2, we activated all the features and asked them to complete the tutorial. To eliminate the possibility that the user might have learned the task during the first stage, which could affect the second stage, we split the group into half and asked the first group to begin with mode 1 while the second group began with mode 2 using the AR assistance system.
 
 \subsection{Completion Time}
\begin{figure}[!h]
	\centering
	\includegraphics[width=3.3in]{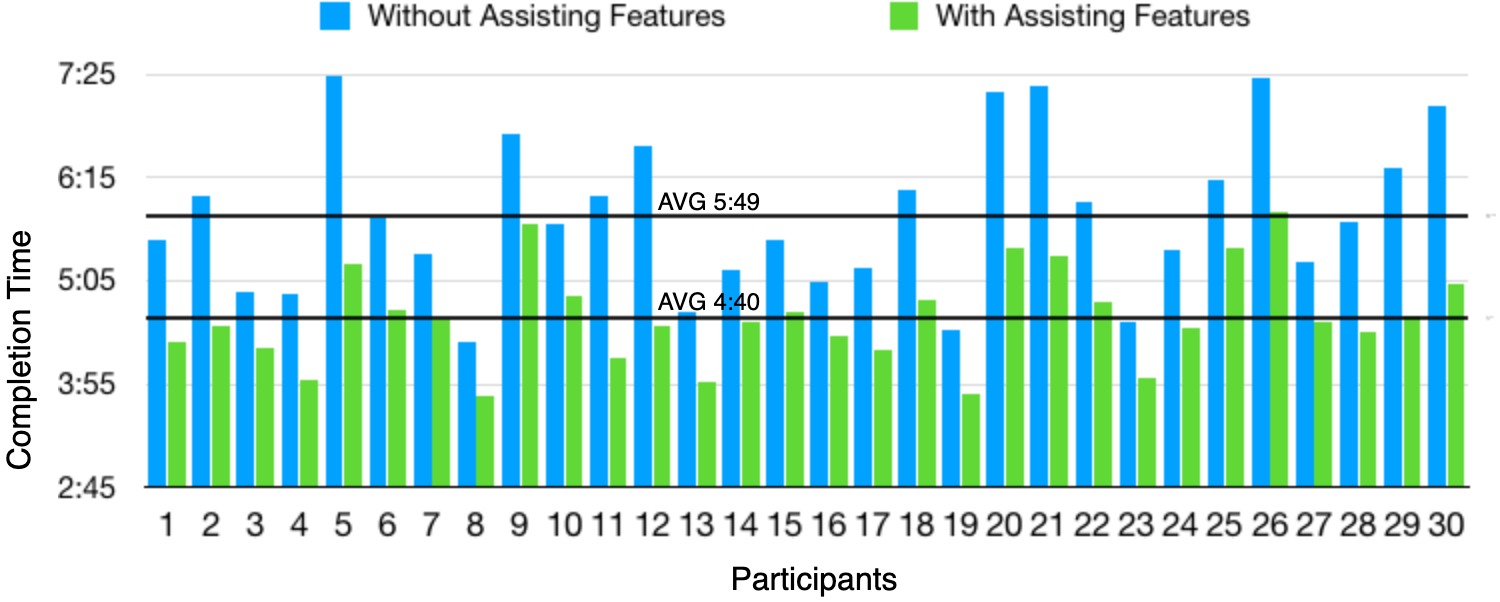}
	\caption{Task completion time}
	\label{time}
\end{figure}

 \noindent First, we evaluated the time difference between the two modes in order to determine if the assistance system leads to a reduced time to task completion. The results are captured in Fig \ref{time}.
\noindent It can be observed that a substantial execution time improvement was achieved when using our system. In every case an improvement occurred. In 10\% of the cases the improvement is over 2 minutes which represents 33\% of saved time. On average, the assembly duration decreases by 69 seconds or 20\%. As the scenario was relatively simple and does not require highly specialized skills, we can expect even better results in the case of very complex scenarios. It is worth noting that during the user study, people with slightly higher technical inclinations performed better (can be noted in participant 3, 4, 8, 19 and 23), requiring less time to complete the tutorial compared to people who have other professional backgrounds where the use of GUI elements decreased the assembly time (can be noted in participant 5, 20, 21 and 28).

\subsection{Error Rate}
\begin{figure}[!h]
	\centering
	\includegraphics[width=3.3in]{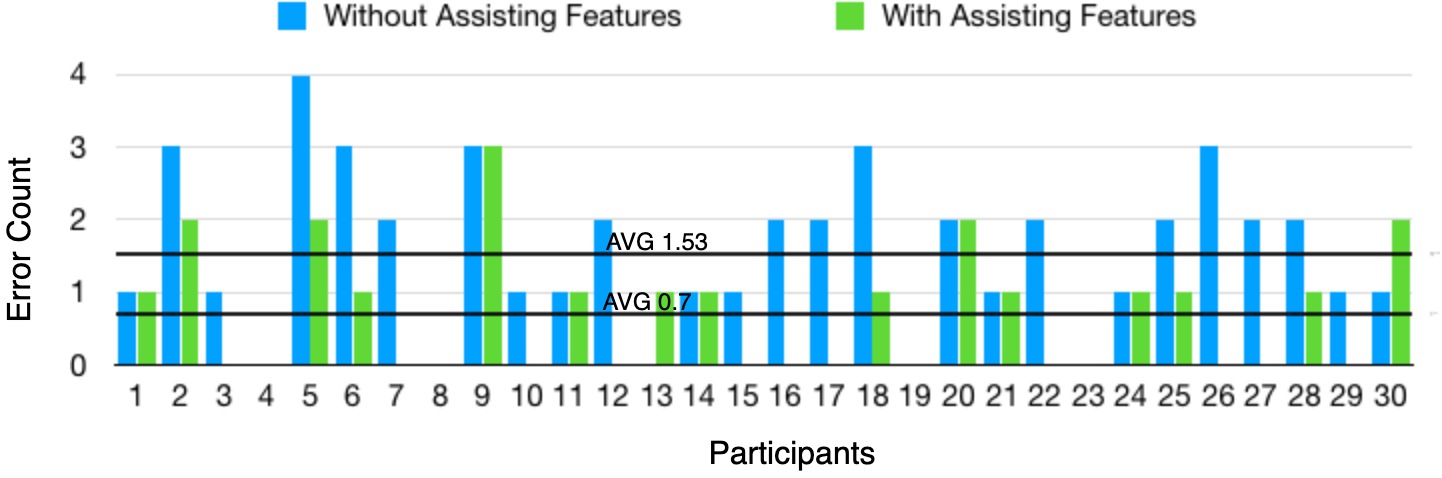}
	\caption{Error Rate}
	\label{error}
\end{figure}

\noindent In the next step we tracked the number of errors the user did on each iteration while doing the above experiment and compared the two modes against each other.
The results are illustrated in Fig. \ref{error}.
The following instances were considered to be errors: grabbing the wrong tool, missing a tool on the working station, performing a wrong action, moving to the next step although the current task was not fully completed and placing an object in the wrong position. Only three users managed to complete the tutorial scenario without any errors on the first try. The rest (90\%) made at least one error. The maximum number of errors registered per person was participant 4. 
During the experimental phase, it was noted that after committing their first mistake users became more alert and started paying more attention to the instructions in an effort to avoid further mistakes. While the average error rate is already small during mode 1 (without AR assistance) with an average of 1.53 errors per user, we still notice a substantial improvement by activating all features in mode\,2. Furthermore, one third of users managed to complete the tutorial without any errors during mode 2. 

\subsection{One-way ANOVA}
\noindent We performed a one-way Analysis of variance (ANOVA) on the data for task completion time (Fig.\,\ref{time}) and error rate (Fig.\,\ref{error}). For the completion time, we found a highly significant difference between the two groups (p=4.83E-8), therefore rejecting the null hypothesis that both groups were randomly sampled from the same distribution. For the error rate, we found a highly significant difference between the groups as well (p=9.3E-4).
The statistical analysis provides strong evidence that our assistance system reduces completion time and reduces error rate significantly in untrained workers when performing the given task for the first time.

\subsection{User Acceptance}
\noindent We evaluated the impact of different components of the assistance system by masking out specific components during the experiments.

\begin{figure}[!h]
	\centering
	\includegraphics[width=3.2in]{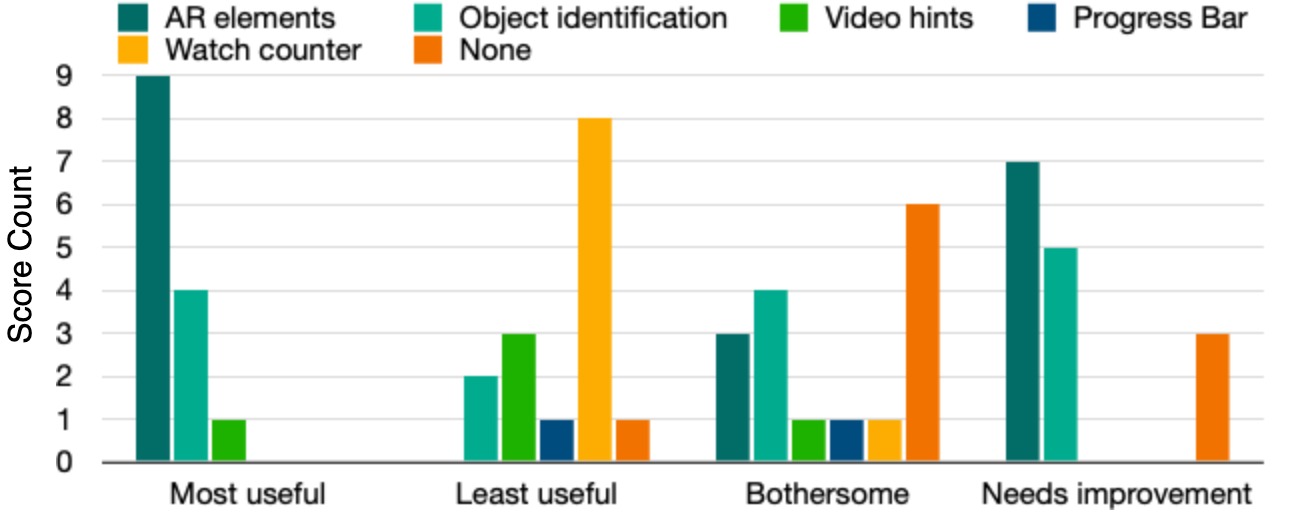}
	\caption{User acceptance of specific AR components.}
	\label{components}
\end{figure}

\noindent As depicted in Fig. \ref{components}, AR elements are considered to be the most useful feature of the GUI. The most popular arguments of the participants included phrases such as "offers extra info", "requires less thinking" and "easy to understand". This indicates that the user acceptance to receive such visual cues is strong. Interestingly, a correlation between "most useful" and "needs improvement" can be observed. Because users put importance on certain features, they also demand that these features work highly accurate. The top arguments for the first two results in the "needs improvement" column were "sometimes elements are slightly off" and "bounding boxes are not always precise". The watch counter was elected as the least useful feature with over 50\% of the votes. Users described it as "not providing any value", "not noticeable" and "does not help with the assembly". Such feedback leads us to believe that the watch counter as a metric should not be used by the worker himself, but rather by managers in order to assess the efficiency of a worker and judge his progress. In the "Bothersome" category, 40\% of the questioned users did not find any of the features to be irritating, which shows that the implemented features complement each other and create an effective learning environment. On the other hand, AR elements and object identification were reported as being "too obvious" and "not necessary". However, we believe that these arguments might be influenced by the relatively low complexity of the tutorial scenario and the common tools used therein. We expect such features to be more valuable in more complex and unknown scenarios.

\subsection{Learning Curve}
\begin{figure}[!h]
	\centering
	\includegraphics[width=3.2in]{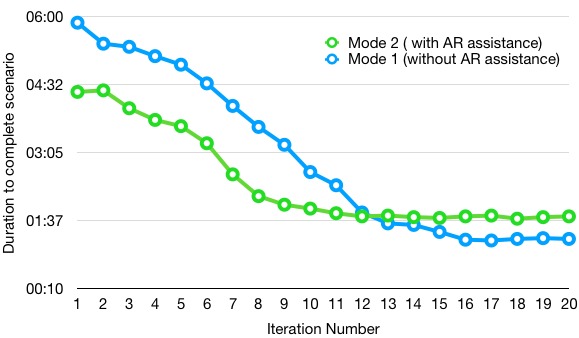}
	\caption{Learning curves of assistance system (blue) and text based instructions only (green)}
	\label{Learning promotion}
\end{figure}

\noindent Lastly, in order to assess the users learning curve, we picked 10 participants out of the original group and asked them to do the same scenario 20 times with and without our assistance system. We then measured each time to complete the task. The results are illustrated in Fig. \ref{Learning promotion}. The completion time converges at the $14^{th}$ iteration while using mode 1 (without AR assistance) and at the $10^{th}$ iteration in mode 2 (with AR assistance). 
Notably, without the usage of the AR assistance users achieve faster task completion after convergence was reached, with the best result requiring only 1:13 minutes while the shortest completion time when using the assistance system is 1:40 minutes. This indicates that the assistance system can help workers learn new workflows faster but decreases the efficiency when the user is already familiar with the task. It is evident that after familiarization with the tasks, completing without the assistance system results in faster completion time as the additional features might become distracting. Furthermore, the scenario was relatively simple. In order to infer more meaningful results, the study has to be extended to more complex tasks in further research. Nevertheless, our study gives empirical evidence that there is a point in time after which AR assistance systems lose their benefits for untrained workers. In our case, this is after 13 iterations of the same task.

\section{CONCLUSION}
\noindent We proposed an AR-based industrial assistance system by incorporating established neural network architectures for computer vision tasks to assist and train workers in manual processes. We combined a robust object detection model with a human pose detection model to recognize the users actions only on RGB camera input and display hints and visual cues accordingly. Furthermore, we integrated and evaluated several AR-based components and additional methods such as video input or animations to assess the impact on user experience and understanding.
The system was evaluated in a variety of relevant metrics for industrial applications such as time to task completion, error rate, user acceptance and learning curve. We performed a one-way ANOVA and found a highly significant (p<0.001) improvement in time efficiency and error rate.
Furthermore, the results indicated a promotion of the learning curve especially for new employees and was highly accepted by participants in the early stages. However, we observed a negative impact with further iterations of the task. This might be attributed to our relatively simple tasks which we conducted for demonstrative purposes. We aspire to extend the user studies to include more complex scenarios and participant groups. Moreover, we aim to add more sophisticated architectures and functionalities to the system to enhance user understanding even further. 
\addtolength{\textheight}{-2cm}   





\bibliographystyle{IEEEtran}

\bibliography{references}

\end{document}